# Morphology of soft and slippery contact via fluid drainage

Yumo Wang[1,2] and Joelle Frechette[2,3*]

[1]National Engineering Laboratory for Pipeline Safety, Beijing Key Laboratory of Urban Oil and Gas Distribution Technology, China University of Petroleum, Beijing, 18# Fuxue Road, Changping District, 102249 Beijing, China

[2]Chemical and Biomolecular Engineering Department and [3]Hopkins Extreme Materials Institute, Johns Hopkins University, Baltimore MD 21218 USA.

(Dated Aug 30th, 2018)

## Abstract

The dynamic of contact formation between soft materials immersed in a fluid is accompanied by fluid drainage and elastic deformation. As a result, controlling the coupling between lubrication pressure and elasticity provides strategies to design materials with reversible and dynamic adhesion to wet or flooded surfaces. We characterize the elastic deformation of a soft coating with nanometer-scale roughness as it approaches and contacts a rigid surface in a fluid environment. The lubrication pressure during the approach causes elastic deformation and prevents contact formation. We observe deformation profiles that are drastically different from those observed for elastic half-space when the thickness of the soft coating is comparable to the hydrodynamic radius. In contrast, we show that surface roughness favors fluid drainage without altering the elastic deformation. As a result, the coupling between elasticity and slip (caused by surface roughness) can lead to trapped fluid pockets in the contact region.

[*]Corresponding author: jfrechette@jhu.edu

# Introduction

Soft coatings on rigid substrates mediate lubricated contact between moving surfaces, and are ubiquitous in tribology[1-3], adhesives[4, 5], and biomaterials[6-8]. Often a coating's function is to mediate contact and near contact interactions with another material. Consider, for example, the emerging field of soft robotics, where compliant coatings are designed to make rapid and repetitive contacts to mimic bio-locomotion[9-12]. Both elastic compliance and surface roughness are hallmarks of biological locomotion, and are expected to govern contact formation and adhesion under flooded conditions[13-15]. In fluids lubrication forces can cause elastohydrodynamic deformation prior to contact[16-18]. Elastohydrodynamic deformation of a soft material during sliding reduces friction and causes lift under loading conditions present during the transport of red blood cells[19] or cartilage lubrication[20] for example. Similarly, elastic deformation due to lubrication forces during a normal approach to contact will hinder bonding and adhesion. Normal approach and detachment is present in bio-inspired wet adhesion[21], dynamic surface force measurements[22-24], particle collision[25], and probe-tack studies of adhesives[26].

While elastic compliance can prevent contact formation, real or apparent surface slip (for example due to roughness) has a competing effect and favors fluid drainage[27]. As most soft surfaces bear some degree of surface roughness, the coupling between elasticity and apparent surface slip is particularly technologically relevant. However how the competition between compliance and roughness affects both fluid drainage and the dynamic of contact formation is poorly understood. In addition, the presence of a soft coating (in contrast to an elastic half-space) introduces a mechanical discontinuity at the boundary between the coating and its underlying rigid substrate, which should also affect the deformation profile caused by lubrication forces. Here we investigate how elastic deformation couples with apparent surface slip caused by surface roughness during a dynamic normal approach to contact.

We use interferometry to image the spatiotemporal surface deformation of a curved surface (radius R) covered with compliant coatings that are much thinner, comparable, and thicker than the hydrodynamic radius ($\sqrt{Rh_0}$), where $h_0$ is the initial separation between the surfaces (see Fig.1). We observe that the deformation profile caused by fluid drainage past a stratified material is qualitatively different from that observed during drainage or indentation past an elastic half-space of the same material. For coatings thicknesses comparable to $\sqrt{Rh_0}$, we see for the first time the formation of elastic double dimples (or wimples). By combining our experiments with modeling, we also show that the coupling between surface roughness and elasticity during contact formation leads to trapped fluid pockets that can be larger than the scale of the surface asperities. Throughout, experiments are validated with a full elastohydrodynamic theory that incorporates the effect of apparent surface slip caused by roughness and stratification. Our measurements when combined with modeling are the first to characterize how roughness affects the drainage process past a wettable and deformable surface.

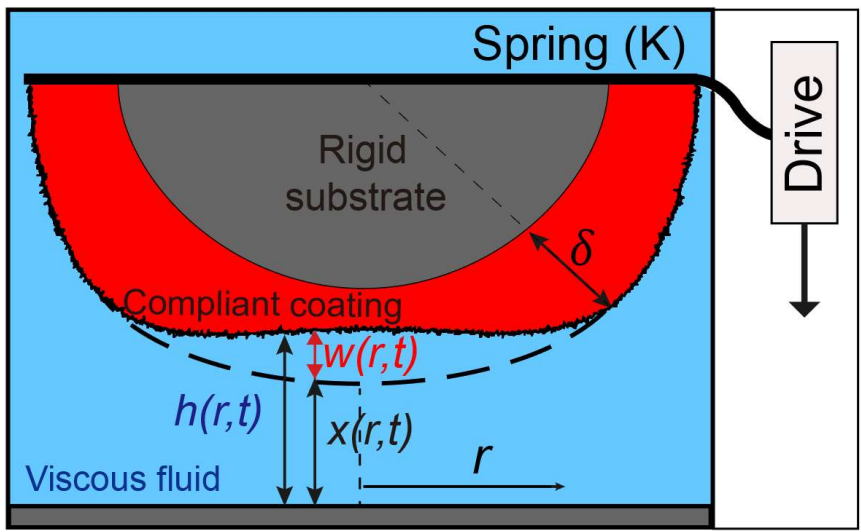

**Figure 1.** Schematics of the experimental configuration with labeled variables (not to scale). Experiments are performed between crossed cylinders, which is equivalent to the sphere-plane geometry shown in the schematic.

## Experimental Details

**Surface Forces Apparatus (SFA)**

We measure the spatiotemporal fluid film thickness in the gap formed between a rigid and a deformable surface (see Fig.1). Our experiments are performed between crossed-cylinders (equivalent to the sphere-plane geometry when $R \gg h$) using the Surface Forces Apparatus (SFA)[28-31]. The deformable surfaces consist of polydimethylsiloxane (PDMS) films of varying thicknesses with a thermally evaporated thin silver layer (50 $nm$) on the top and glued, silver side up, on a SFA cylinder. The silver layer on top of the PDMS serves a mirror for the interferometer, prevents swelling of the PDMS with the silicone oil fluid ($\eta = 0.2\, pa \cdot s$), and introduces surface roughness from thermal evaporation[32]. We rely on white light interferometry[29, 33, 34] to map the absolute local fluid film thickness, $h(r,t)$ with nanometer resolution in the normal direction and micron resolution in the lateral direction.

Initially, the interacting surfaces are at rest and separated by $h_0 \sim$ 2-5 $\mu m$. The measurements are initiated with the motor connected to the lower rigid surface moving at a constant drive velocity $V$ (between 50-150 $nm/s$). The cantilever spring ($k = 165.31\, N/m$) connecting the motor and lower surface deflects due to the viscous forces, therefore the approaching speed of the two surfaces is always less than $V$. We determine independently the Young's modulus of the PDMS (~1*MPa*) via indentation measurements with a multifunctional force microscope[35]. We validate our data using elastohydrodynamic theory[36] in which we incorporate corrections for apparent surface slip in the form of a shifted plane[37]. As references for limiting cases we show Reynolds' theory for rigid surfaces[38] and Davis-Serayssol-Hinch (DSH) theory for elastic half-space[39]. For further details of data processing and analysis see section 3 of Supplemental Material.

**Experimental Methods**

An overview of the experimental parameters investigated in this work is presented in Table S1 in Supplementary Material. The PDMS coating thickness is measured through laser profilometry (Filmetrics F20-NIR) and verified by digital calipers. We rely on white light multiple-beam interferometry, with silver films as reflective mirrors to map directly the local fluid film thickness. Note that we use the center of mass of the interferometric fringes to calculate the fluid film thickness. The fringes represent the summation of all intensity profiles corresponding to the point of lights transmitted through silver mirrors. As a result, the fluid film thickness we measured is based on the center of mass of the asperities within each pixel (~2 *μm*). The disks radii are measured during the experiments from multiple-beam interferometry at rest.

The contact position (zero separation) is defined through quasi-static experiments in which the surfaces were slowly pushed together until a large central contact regime is formed (See Figure S6 in Supporting Info). We mark the position at the center pixel (point of closest approach) and use it as the reference position for zero separation throughout. The lateral variation of this contact position is typically less than 0.5 *nm* (Figure S6A in Supporting Info) and can be attributed to the roughness of the silver layer. For some cases (Figure S6B in Supporting Info) the tilting or unevenness of underlying SU-8 can affect the contact morphology in which the maximum spacial variation can be 5 *nm*.

The bulk Young's modulus for the polydimethylsiloxane, PDMS is measured via indentation with a spherical probe by a home-built Multifunctional Force Microscope[35]. The indentation data is analyzed with a model that corrects for the finite thickness of the PDMS coating[40]. Throughout the experiments, we use silicone oil (Xiameter PMX 200 Silicone Fluid) purchased from Dow Corning without further treatment as the working fluid. For more details regarding surface characterization please see Supplementary Material.

**Fabrication of SU-8 Surface**

We rely on the negative photoresist SU-8 (SU-8 2007, MicroChem) as a coating on a muscovite mica sheet as a smooth and rigid surface. The mica pieces are cleaved to a thickness of 2-8 *μm* using a needle in a laminar flow hood. The freshly cleaved samples are placed on larger and thicker mica backing sheet (one piece each backing sheet). A 50 *nm* silver layer is deposited at a rate of 0.3 *nm/s* on top of the mica sheet using a thermal evaporator (Kurt J. Lesker Nano38). Note that the mica surfaces here will not be a part of the interferometer in the SFA but serve as a support for the SU-8 layer. The silvered backing sheet is then attached to a silicon wafer and the negative photoresist is spin-coated on the mica at 5000 *rpm* for 1 minute, resulting in a thickness of 6-7 *μm*. After spin-coating, the individual mica pieces are transferred to a new backing sheet and baked on a hot plate at 65 $^oC$ -95 $^oC$ -65 $^oC$ for 3-5-3 minutes to follow the supplier's guidelines. The backing sheet is then exposed to UV light using Mask Aligner (EVG 620) at constant energy mode to receive 140 $mJ/m^2$, followed by a hard bake at 150 $^oC$ for 1 min. After baking, an individual mica sheet is peeled from the backing sheet and glued (mica side down) to a SFA cylindrical lens.

**Fabrication of PDMS Coating**

We deposit the PDMS on silicon wafers prior to gluing the PDMS films on the SFA disks. The wafers are pre-treated with trichloro(octadecyl)silane (OTS) to facilitate lift-off after deposition. The silanization process followed vapor deposition protocols described in the literature[41]. We then mix PDMS (Sylgard 184) pre-polymer and the crosslinking agent at a ratio of 10:1 using an automatic mixing instrument (Elveflow). The PDMS is then spin-coated on the treated wafer using different speeds to obtain different thicknesses, followed by curing first at 75$^oC$ for 3 *h* and then left at room temperature overnight. The cured films were then extracted in hexane for 24 *h* and sonicated in ethanol 3 times, 15 minutes each to remove unreacted oligomers. After drying the PDMS in a laminar hood, we cut the film into 1 *cm* x1 *cm* squares using a razor blade and carefully lifted the sheet with a point tweezer and glued to a SFA cylinder. We used the remaining samples to measure the thickness of PDMS film using laser profilometry (Filmetrics F20-NIR) and with digital calipers. The cylinders were then putted into a custom-built plasma reactor for oxygen plasma at 50 *W* under 0.3 *Torr* oxygen for 5 seconds. Longer exposure to the oxygen plasma lead to cracking, shorter or no exposure lead to wavy surface structure and poor adhesion between silver and PDMS[32]. After the plasma treatment, the cylinders were directly putted into a thermal evaporator (Kurt J. Lesker Nano38) and coated with 50 *nm* silver (99.999% purity, Alfa Aesar) at 0.3 *nm/s*. The silvered cylinders were then taken out and assembled into SFA as the top surface. We prepare 4 sets of PDMS film with thickness ranging from 11 *μm* to 330 *μm*, and test separately for each thickness. For further details and explanations regarding surface fabrication, we guide the readers to Ref.[42].

# Overview of theoretical treatment

The theoretical developments and numerical solutions for the drainage past a soft coatings have been studied by several groups[36, 43, 44]. We recently reported an analysis directly applicable to dynamic force measurements[36] that we employ here as is, and also with an additional correction for apparent surface slip. A brief summary of the theoretical treatment is presented here for clarity, a more detailed description of the theoretical development and numerical methods is available in Ref[36].

Consider normal fluid drainage in the presence of a stratified elastic boundary in the sphere-plane geometry. The relative movement of the two surfaces can exert fluid pressures, and cause elastic deformation of the soft layer, which is bound on one of the two surfaces (Fig. 1). The fluid flow is described by the lubrication approximation, which is valid in the limit of low Reynolds number and when the central fluid film thickness, $h(r=0,t)$, is small compared to the sphere radius ( $R$ ), with $r$ being the radial coordinate and $t$ the time. If we assume the no-slip boundary condition on both surfaces and continuum fluid phase, the axisymmetric drainage and infusion of a Newtonian fluid (viscosity $\eta$ ) from a thin gap is given by:

$$\frac{\partial h(r,t)}{\partial t} = \frac{1}{12\eta r}\frac{\partial}{\partial r}\left(rh^3 \frac{\partial p(r,t)}{\partial r}\right), \tag{1}$$

showing that the fluid pressure distribution, $p(r,t)$, between the two surfaces is related to the shape of the gap and to the rate at which the surfaces approach one another. We use the equation as is for the no-slip case. To account for apparent slip caused by surface roughness we also introduce a shifted plane model in the surface separation. Since we calculate absolute surface separation for each time at each radial position, we can simply add the shifted length ( $h_s$ ) to the separation in the lubrication Equation 1. The shifted plane length ( $h_s$ ) is not known *a priori* but used as a fitted parameter using multiple experiments. The lubrication equation can be corrected as:

$$\frac{\partial h(r,t)}{\partial t} = \frac{1}{12\eta r}\frac{\partial}{\partial r}\left(r\left(h+h_s\right)^3 \frac{\partial p(r,t)}{\partial r}\right), \tag{2}$$

We calculate absolute surface separation for each time step and for all radial positions, we then add the shifted length to the separation in the lubrication equation, and keep all other parts of the numerical calculation identical to our previously developed algorithm[36].

We use the linear elasticity theory for stratified materials to describe the mechanical response of the surface coating where the source of surface stress is the fluid pressure distribution. In our present geometry, the radius of the sphere is much greater than the fluid film thickness ($R >> h$), so that the local curvature of spherical surface is very small. In this case, the normal stress $\sigma_N$ dominates over tangential stress $\sigma_T$ at the surface, because $\sigma_T/\sigma_N \sim \sqrt{h/R} << 1$. As a result, we can take the fluid pressure as the normal boundary pressure and neglect the shear stress at the surface. Within the framework of linear elasticity, we obtain the deformation distribution of an elastic layer by solving the biharmonic equation, which can be reduced to an ODE by using Hankel transforms in cylindrical coordinates.

The four boundary conditions in our configuration are an axisymmetric normal stress and a negligible shear stress on top of the soft coatings, together with sticky boundary conditions (zero normal and shear deformation) at the bottom of the substrate-coating surface. For this specific case, a closed form for the surface deformation, which is needed to calculate $h(r,t)$ in Equation 2 was derived previously in the context of indentation[45, 46] and used by others in the context of elastohydrodynamics[43, 47]. Here, the surface deformation can be calculated from:

$$w(r) = \int_0^\infty \frac{2}{E^*\xi} X(\xi\delta) Z(\xi) J_0(\xi r) d\xi, \tag{3}$$

where

$$X(\xi\delta) = \frac{\gamma(1-e^{-4\xi\delta}) - 4\xi\delta e^{-2\xi\delta}}{\gamma(1+e^{-4\xi\delta}) + (\gamma^2 + 1 + 4(\xi\delta)^2)e^{-2\xi\delta}} \; ; \; \gamma = 3 - 4\nu \tag{4}$$

and

$$Z(\xi) = \xi \int_0^\infty r p(r) J_0(\xi r) dr . \tag{5}$$

In Equations 3-5, $p(r,t)$ represent the applied normal traction, which is the local liquid pressure on the surface at a given radial position $r$. $Z(\xi)$ is the modified Hankel transform of the pressure, in which $J_0(\xi r)$ is the 0th-order Bessel function of the first kind. $\delta$ is the compliant film thickness, $\nu$ is Poisson's ratio of the soft coating, and $E^*$ is the reduced Young's modulus.

During the experiments, one surface is mounted on a cantilever spring. In this case, the surface separation is different from the displacement of the driving motor due to the deflection of the spring. Another equation describing the mechanical coupling between the spring and hydrodynamic forces is necessary to describe the drainage process. As shown in the following equation:

$$F_s = kS(t) = k(h - h_0 + Vt - w) = F_H = \int_0^R 2\pi r p(r) dr \tag{6}$$

$F_s$ is the spring force which is balanced by the hydrodynamic force, $F_H$. The spring force is a product of the spring constant, $k$, and the deflection, $S(t)$, which is calculated from the initial separation at the centerpoint, $h_0 = h(0,0)$.

To simplify our calculation, we rescaled the equations above to make them dimensionless. The resulting dimensionless parameters are:

$$\hat{r} = \frac{r}{\sqrt{Rh_0}}, \quad \hat{h} = h/h_0$$
$$\hat{t} = \frac{V}{h_0}t, \quad \hat{p} = \frac{h_0^2}{\eta R V}p \quad (7)$$
$$\hat{w} = \frac{w}{h_0}, \quad T = \delta/\sqrt{Rh_0}$$

We solve Equations 2-7 simultaneously using MATLAB to obtain the spatiotemporal surface deformation, viscous forces, fluid pressure, and fluid film thickness during experiments. We only consider the central part of the spherical surface (for $r < 0.1R$) where the drainage pressure is the largest, and the pressure past $r < 0.1R$ is set at 0 because it is negligible compared to the central pressure. The central region ($r < 0.1R$) is discretized into 500 evenly spaced points and the special and temporal derivatives of variables are obtained using finite difference method. We start the calculation from the initial separation at $\hat{t} = 0$ when the motor starts moving (from rest) at a constant drive velocity. An iteration algorithm that is similar to the one employed in Ref.[39] is incorporated for calculating the fluid film thickness at each time. In short, the surface separation $h(r)$ is first estimated based on results from the previous time step, then iterated from Equations 2,3 and 6 until the maximum error between calculated $h(r)$ and estimated $h(r)$ less than 0.01 *nm*, before we proceed to next time step.

We also perform a convergence test for all discretized variables, including radial position and time. The mesh size on both $\hat{t}$ and $\hat{r}$ are decreased until the change in the calculated hydrodynamic force between subsequent iterations is less than 1%. In all of our results, the increment in dimensionless time is set as 0.00667, and the mesh size in dimensionless radial position is 0.0163. We use Gauss–Legendre quadrature method to calculate the integrals.

We validate our model by recovering two known limits: the Reynolds' theory for rigid surfaces[38] and DSH (Davis–Seyrassol–Hinch) model between elastic half-space[39]. We simply change the coating thickness to extremely small or large values, without different assumptions. The results show excellent agreements between our model and analytical equations for limiting cases[36].

We do not expect the 50 *nm* silver film on the PDMS to have significant constraint on the deformation of PDMS layer, because of the thickness of the silver film compared to that of the PDMS (50 *nm* vs 10-330 microns) along with the large radius of curvature in the SFA (1-2 *cm*). We therefore neglect the silver layer in our theoretical development. Detailed verification showing that the silver layer can be neglected can be found in Supplementary Information with knowledge from Refs. [48, 49].

## Results

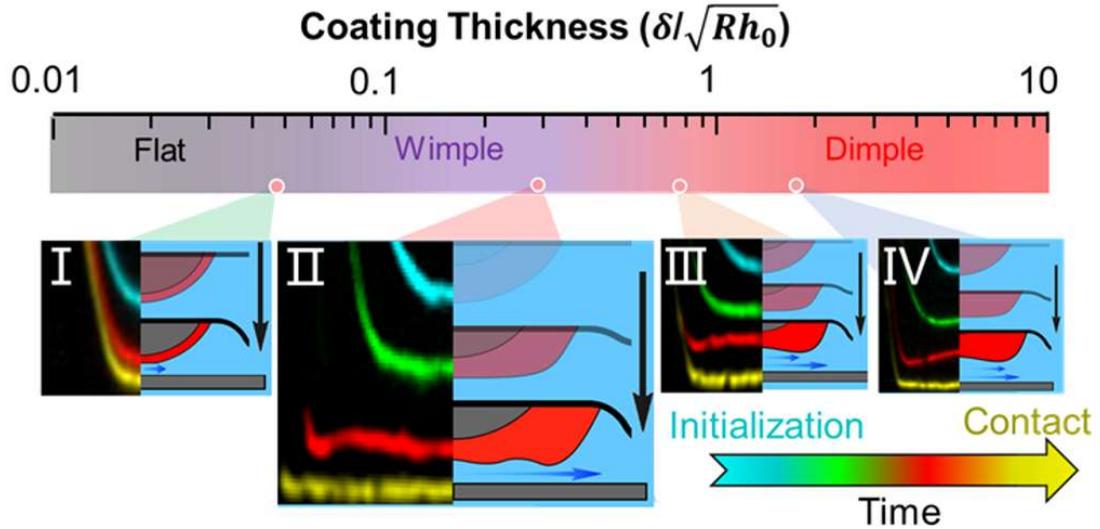

**Figure 2.** Surface morphology of soft coatings of different thicknesses during normal fluid drainage. Deformation profiles are shown from I to IV for increasing coating thicknesses, $T = \delta / \sqrt{Rh_0}$. The coating thickness is illustrated in the line bar on top. For (I-IV) on the bottom the left sides show pictures of individual interferometric fringes captured by SFA experiments, with different colors representing timepoints during drainage. The color scheme is labeled by the arrow on bottom. The yellow fringe represents the contact position from quasi-static experiments. The right sides of (I-IV) show schematics illustrating the process (not to scale).

**Observation of wimples**

The effect of the coating thickness on the elastohydrodynamic deformation during the drainage process is clearly shown in Fig.2. For thick films a dimple forms at the center where the fluid pressure is highest (case IV in Fig.2). For coatings that have a thickness comparable to the hydrodynamic radius, we observe the formation of a second dimple close to contact (case II in Fig.2). Similar double dimples morphologies have been observed during drainage past fluid droplets, and have been named "wimples"[50]. However, there are no prior reports of this type of deformation for elastic solids. The formation of wimples here is a direct consequence of the combination of stratification and surface curvature and has a distinct origin than for fluid droplets. For a half-space, the fluid pressure distribution is highest at the centerpoint and the associated surface stress can be fully accommodated by normal deformation within the coating, thus a dimple forms[51]. For a film with a finite thickness, as shown by the case II in Fig. 2, the fluid pressure remains the largest at center, and the deformation is still largest at the center (Fig.3a). However, the deformation is hindered in the central region because of the mechanical constraints imposed by the supporting rigid substrate, leading to broadening of the deformation profile compared to that of a half-space. A wimple is observed when the radius of the deformation profile becomes comparable with the original surface curvature, and results from the superimposition of radial deformation profile on a curved surface (the deformation profile is still monotonic). Counterintuitively, during the drainage process the outer dimple is first visible, followed by the central one (Fig.3a).

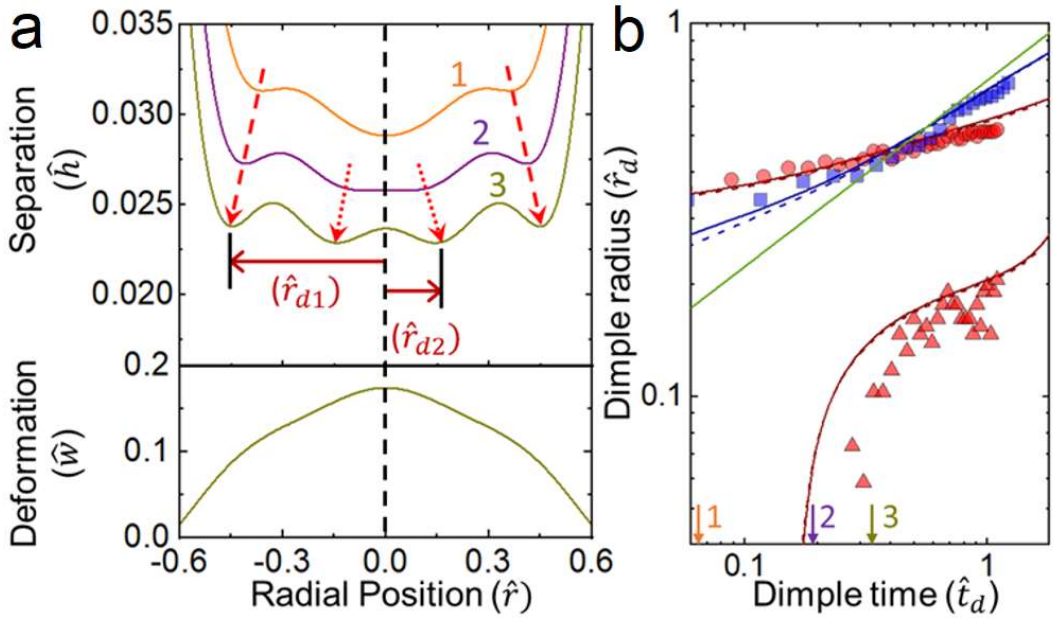

**Figure 3.** (a) (Top) Predictions of the surface morphology (dimensionless fluid film thickness: $\hat{h} = h/h_o$) when the wimple forms for case (II) given in Figure 2. The number identifying each curve correspond to the timepoints indicated by arrows in (**b**). Red dashed arrows show the primary dimple growth and the dotted arrows show the secondary dimple growth. The primary and secondary dimple radii, $\hat{r}_{d1}$ and $\hat{r}_{d2}$, are defined by the solid red arrows. (Bottom) Dimensionless surface deformation, $\hat{w} = h/h_o$ for case (II), at timepoint 3. (**b**) Dimensionless dimple radius ($r_d / \sqrt{Rh_0}$) as a function of dimensionless dimple formation time ($t_d / (h_0/V)$), $\hat{t}_d = 0$ when a dimple is first observed. Blue square points and line: approximated half-space data ($T = 1.6$) and theory (case IV in Figure 2). The case II in Figure 2 ($T = 0.3$) is shown in red. The circle points show the time evolution of the $\hat{r}_{d1}$ and the triangle points are for $\hat{r}_{d2}$. Solid lines: predictions for a stratified material with a 35nm shifted length, dashed lines: stratification and a no-slip boundary condition. The green line represents the scaling of $\hat{r}_d = \sqrt{\hat{t}_d / 2}$.

Wimples can be predicted for all thicknesses below the half-space limit. Although practically they are difficult to visualize in experiments if the normalized coating thickness, $T = \delta / \sqrt{Rh_0}$, is not 0.1-1. For thinner coatings, the small amplitude of the wimples can limit experimental detection, and would occur at separations where other effects such as surface forces could also affect the surface profile[27, 52]. Based on these considerations, a wimple is most likely to be visible (normal amplitude ≥ 5 nm, SFA resolution < 1nm) where $\delta / \sqrt{Rh_0}$ ~ 0.1-0.3. In our experiments, the wimple amplitude is typically ~ 10 nm, while dimples can be as large as 50 nm[51]. For colloidal probe AFM force measurements, wimples could occur for a probe of R = 100 μm with E~1MPa coating ($\delta$ = 2 μm) with an initial position $h_0$ ~ 500 nm with V ~ 2 μm/s. Note that we use a viscous silicone oil fluid. For experiments in an aqueous solution a lower modulus of the coating is necessary to recover deformation of similar amplitudes.

The dynamics of wimple growth is also quite different from that of dimples. In Fig.3b, we show the growth rate of the radius of a primary and secondary dimple (defined in Fig.3a) for $\delta$ = 54 μm ($T = \delta / \sqrt{Rh_0}$ ~0.3, case II in Fig.2), along with predictions from our elastohydrodynamic model. The dimple radius $\hat{r}_d$ is non-dimensionalized by $\sqrt{Rh_0}$, while time is normalized by $h_0 / V$, and $\hat{t}_d = 0$ when a dimple is first observed. For $T$ = 0.3, initially only the primary dimple is visible (orange line in Fig.3a), meaning that the initial dimple starts to grow from the edge rather than grow from the center (as would be the case for an elastic half-space). Once the secondary dimple forms, both radii increase with time, suggesting a continuous broadening of the central deformed region. Our model shows excellent agreement with the measurements, but we slightly overestimate both radii (Fig.3b). We also plot the geometrical scaling of $\hat{r}_d = \sqrt{\hat{t}_d / 2}$ in green line, which is obtained from volume conservation between half-spaces, and is routinely employed for contact mechanics of thick films and for droplets impact[51, 53, 54]. We see that even for a thick coating (blue), the scaling of $\hat{r}_d = \sqrt{\hat{t}_d / 2}$ can be used only at long times, and for thinner coatings the growth rate of $\hat{r}_d$ is much slower than that of the half-space, indicating that the normal displacement of the soft coating is hampered by the rigid substrate.

**Importance of hydrodynamic boundary condition during drainage**

Drainage curves, where the fluid film thickness is measured as a function of time, or the measurement of lubrication forces are routinely employed to characterize the hydrodynamic boundary conditions of a solid-liquid interfaces[55, 56]. Typical drainage curves displaying the central separation, $h(0,t)$, as a function of time are shown in Fig.4 for two coating thicknesses: $\delta$=147μm, $\delta/\sqrt{Rh_0}$ ~ 0.8 (Fig.4a), and $\delta$=54μm, $\delta/\sqrt{Rh_0}$ ~ 0.3 (Fig.4b). We also plot the hydrodynamic force as a function of $1/h$ to highlight the importance of the hydrodynamic boundary condition at short range (inset). Predictions for the drainage curves representing different hydrodynamic boundary conditions are also plotted and discussed below. While the two measurements are qualitatively similar, a wimple is present starting at $t = 65.0\ s$ in the $\delta=54\ \mu m$ coating, but not in the $\delta=147\ \mu m$ one. More generally, the formation of a wimple cannot be detected from visual inspection of the drainage or force curves alone.

For both coatings in Fig.4 we observe significant decelerations as fluid drains out of the gap, ultimately reaching a limit where the drainage rate becomes insignificant, and a finite fluid film is trapped at the center (contact, $h=0$, is not reached). Surface forces [51, 57] could favor contact formation, but for both cases shown in Fig.4, the central separations at long times are larger than the typical range of these interactions. In contrast, under the same conditions rigid surfaces would likely have made contact due to surface forces[58], see the predictions from Reynolds theory shown as dashed lines. When comparing the effect of the thickness of the PDMS coating on the drainage curves, we see that for a thicker coating (Fig.4a) more fluid is trapped at the center at long times (and hydrodynamic forces are larger at a given separation). For a thinner elastic coating, stratification constrains more the normal deformation and, as a result, more energy can be dissipated in fluid drainage rather than in elastic deformation, leading to smaller separations between the two surfaces.

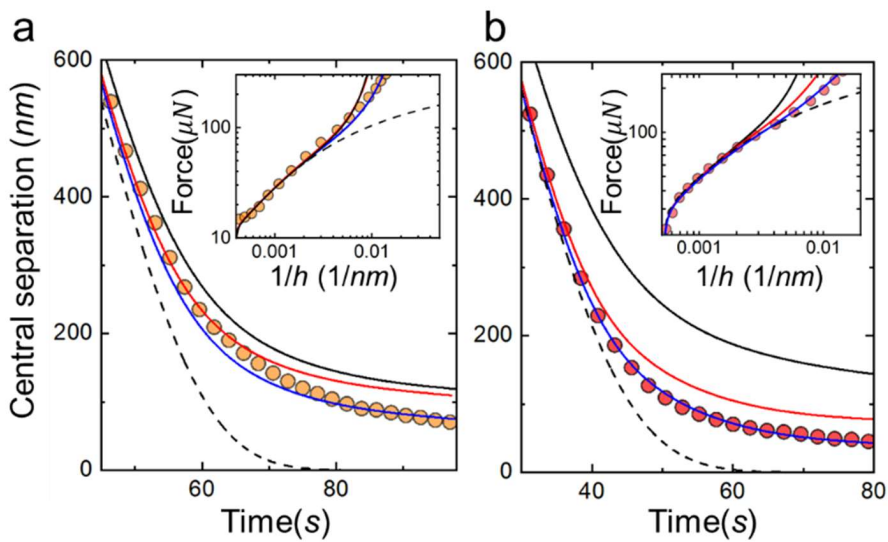

**Figure 4.** Importance of the hydrodynamic boundary condition during fluid drainage. Central fluid film thickness between the approaching surfaces shown for two coating thicknesses (**a**) $T = 0.8$, $h_0$=2558 $nm$, $V$ = 52 $nm/s$ and (**b**) $T = 0.3$, $h_0$=1974 $nm$, $V$ = 62 $nm/s$. Lines are predictions for half space (black), stratified theory with no-slip boundary condition (red), stratified theory combined with a shifted plane correction of 35 $nm$ (blue), and Reynolds theory for rigid surfaces (dashed). Inset: Repulsive hydrodynamic force as a function of the inversed central separation.

The role of the hydrodynamic boundary condition is clearly visible from the solid lines in Fig.4, where we present a systematic and gradual increase in complexity. In our prior work we showed that for thick coatings ($T$~1.6), introducing an elastic boundary condition that corresponds to a half-space could describe data well[51]. For thinner coatings, such as the ones shown in Fig.4, we now see that an elastic boundary condition (black lines) overestimates the measured fluid film thickness and the hydrodynamic forces. We also see that the departure from the half-space limit increases as the coating thickness decreases: the data in Fig.4a is closer to the half-space limit than the data in Fig.4b. Correcting for the finite coating thickness using a no-slip stratified boundary condition[36] improves the agreement with experiments (red lines), especially for the $T$~0.3 coating (Fig. 4b). For both coating thicknesses, predictions for drainage past a stratified material describes the data better than either limiting cases (rigid or half-space, black solid and dashed lines). However, at longer times when the surfaces are closer to contact, the no-slip stratified boundary condition also systematically overestimates the measured data, demonstrating that surface roughness competes with elastic deformation and facilitates drainage. The effect of roughness is observed for fluid gaps as large as 300 *nm*. In this regime, a stratified boundary with a shifted no-slip plane is necessary to account for surface roughness, and it captures the data very well for both coatings (blue line). In fact, at all times, the stratified elastohydrodynamic theory with a shifted length (blue line), which is a combination of surface compliance, stratification, and roughness, best describes the experimental data for the whole drainage process. As both the surface roughness and stratification facilitate drainage, these two effects could be indistinguishable from measured force curves and lead to misinterpretation during dynamic force measurements.

We suspect that the correction for apparent slip necessary to capture the data at short range in Fig.4 are caused by the effect of surface roughness on the no-slip boundary conditions at the silver-liquid interface (Fig.1). Hydrodynamic forces exhibits different asymptotic behavior with decreasing surface separation during the drainage past rigid wettable surfaces with nanometer scale roughness, leading to weaker forces than can be predicted by the no-slip boundary condition[37]. A more realistic boundary condition for rough wettable surfaces would be that of no-slip, but with the no-slip plane located between the top and bottom of the asperities[37, 59, 60]. This correction is distinct from that of pure slip, in that the shift due to roughness is independent of the fluid film thickness[60, 61]. Here we introduce a shifted plane model that we attribute to the presence of surface roughness. Incorporating this correction for a stratified deformable surface is more cumbersome than for rigid surface[62]. The shifted length is not known *a priori* but used as a fitted parameter. We use the drainage data from multiple experiments (velocity and coating thickness) and select a single shifted plane value of 35 *nm* that minimize the overall error and gives the best agreement for the drainage curves (blue lines in Fig.4). A better agreement could be achieved if a different shifted plane value was employed for different experiments. Similarly, for simplicity we assume that the shifted plane is constant at all drive velocities, even if reports suggest a possible dependence between shifted length and the drive velocity when the surface separation is of order of the surface roughness[5]. This value for the shifted plane is comparable to the peak-to-valley roughness of the silver film on the PDMS coating obtained independently. We scanned on an area of $3 \times 3$ $\mu m^2$ and the RMS roughness was between 3-6 $nm$, while the peak-to-valley roughness was between 20-50 $nm$, depending on experiments and samples. When incorporating the shifted plane, the elastohydrodynamic model describes accurately our measurements. To the best of our knowledge these are the first measurements characterizing the combined effect of surface roughness and elasticity on fluid drainage.

When simply looking at the force or drainage curves, it is difficult to distinguish the roles played by surface roughness or elasticity without prior knowledge of the film thickness and mechanical properties (Fig.4). However, we find that the contributions of stratification and slip on the drainage process are clearly distinct and can be decoupled more readily when observing the surface morphology, as shown in Fig.5 for the same two coating thicknesses of Fig.4. When predicting the spatiotemporal fluid film profiles, including the wimples observed in the experiments, we find that the morphology is almost identical whether or not slip is incorporated in the elastohydrodynamic theory for stratified materials[36], see the red and blue lines in Fig.5. In fact, the morphology is almost uniquely determined by the properties of the compliant coating (modulus and thickness), whereas the shifted plane model shifts the position of the surface without altering the morphology significantly (for comparison between different shifted lengths see section 2.3 in Supplementary Material). For $r < 80$ $\mu m$, the maximum error between data and theory is 8 $nm$. The theory for half-space predicts a single dimple at the center, showing qualitative difference from the profile observed.

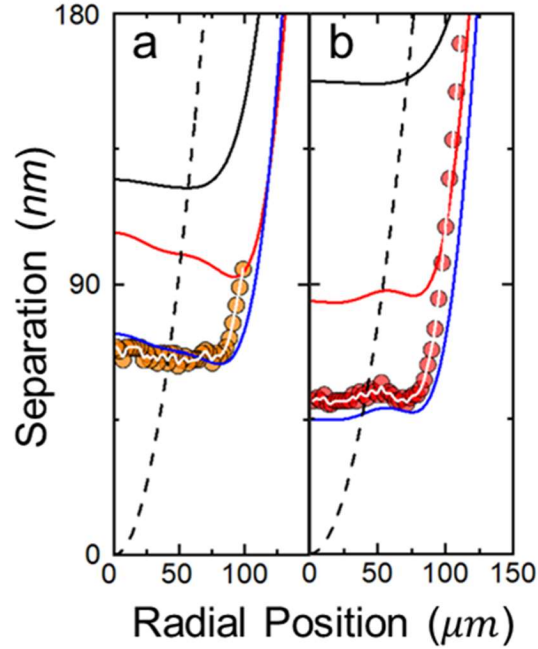

**Figure 5.** Role of hydrodynamic boundary condition on surface morphology. (**a**) $\delta = 147\,\mu m$, at $t = 95.2\,s$ in Fig.4a. (**b**) $\delta = 54\,\mu m$, at $t = 71.2\,s$ in Fig.4b. Data points are the experimental results, the white lines are to guide the eyes. Predictions for stratified material with (red) no-slip and (blue) combined with a shifted plane correction of 35 $nm$ (blue). Other lines are for half space (black) and rigid surfaces (dashed).

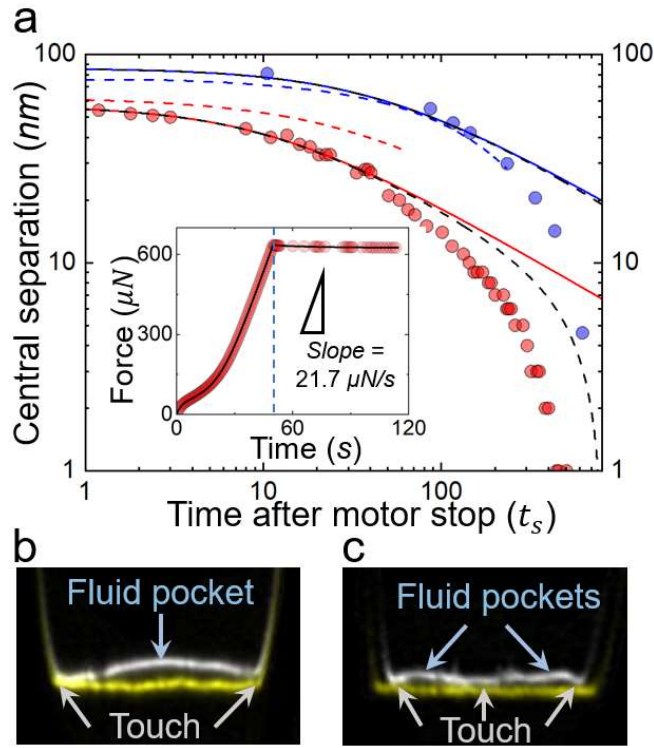

**Figure 6.** Trapped fluid pockets between surfaces in contact. (**a**) Relaxation at constant force at the center point after the motor stops ($t_s$). Points are experimental results and lines are prediction for the Stefan-Reynolds model with $\tau$ : 90.0s and 22.4s for blue and red curves, respectively. Colored dashed lines: stratified model with slip length 35 *nm*, *T* = 1.6 for blue, and *T* = 0.3 for red. Black dashed lines: Stefan-Reynolds model including the van der Waals interaction (use Hamaker constant $4 \times 10^{-19}$ *J*). Inset: Force versus time where the motor stop at *t* = 50*s* (dashed line) after which the force is constant. (**b-c**) Interferometric fringes showing contact morphology for *T*=1.6 (**b**) and *T*=0.3 (**c**).

**Trapped fluid pockets in contact**

The importance of surface roughness and the limitations of the shifted plane correction are clearly visible during constant force drainage, which is achieved by stopping the motor once the spring is deflected (inset, Fig.6a). For constant force drainage the time-dependent fluid film thickness at the centerpoint can be approximated by the Stefan-Reynolds equation $h/h_{stop} = (1 + 2t_s/\tau)^{-0.5}$ [38, 63], through a fitted time scale $\tau$ [64], and ignoring surface forces. Here $t_s$ and $h_{stop} \sim 50 - 500\,nm$ are the time and central separation when the motor stops, respectively. We compare this approximation with our data, and with the predictions for drainage past a stratified material that include the shifted plane (dashed lines). We see that for *h* > 20-30 *nm* the approximation describes very well the drainage process, and then overestimates the fluid film thickness at shorter separation. We also see that the theory that includes the shifted plane deviates from the surface separation in the region where the fluid film thickness is of order of the surface roughness. This discrepancy with experiments is likely due to roughness and limitations of the shifted plane approximation in the thin channel limit[62, 65, 66], where the surface separation is comparable to the length scale characteristic of the surface roughness. As the surfaces are close to contact, the approximation of treating the rough surface as a flat homogenous surface with an averaged boundary is unlikely to be correct. Such deviations have been shown for rigid surfaces and are observed here for a compliant coating and point to the need of a better hydrodynamic boundary condition in this limit. Note here that the deviations are unlikely due to surface forces, because the two surfaces are still ~ 50 *nm* away when the theory starts to deviate. In fact, as an absolute upper bound, if we estimate the van der Waals interactions between two surfaces by taking the Hamaker constant for Silver-Air-Silver ($\sim 4 \times 10^{-19}$ *J*), the attractive interaction cannot exceed 3 $\mu N$ at a separation of 20 *nm*, which is negligible compared to repulsive hydrodynamic force we measure at that separation (Inset of Fig.6a). We expect the van der Waals interactions to be significantly weaker than this estimate because (1) the silver film is very thin (50*nm*) and not a half-space, (2) the second opposing surface is a smooth SU-8 layer, which has a much lower surface energy, see Figure S1 in Supplementary Information (3) the intervening medium is a silicon oil with a refractive index similar to that of PDMS and SU-8. However even an overestimation of van der Waals interactions show that an attraction between the surfaces is not sufficient to explain the discrepancy between the experiments and the predictions (Black dashed lines in Fig.6a, using Eqns. 46-47 in Ref.[64]).

The constant force drainage experiments allow us to monitor contact formation. Once the motor stops, contact is reached after a long relaxation. Roughness facilitates drainage through asperities and effectively competes with elastohydrodynamic deformation. Interestingly, although a long relaxation time is needed, the dimple or wimple remains essentially intact, even upon boundary contact (Fig.6b-c). The image of the initial contact morphology is shown in white ($T \sim 1.6$, $t_s = 2.52$ min for Fig. 6b; $T \sim 0.3$, $t_s = 2.75$ min for Fig.6c). The yellow-colored fringes indicate the measured final contact formation obtained after a long relaxation (> 10 min) at constant force, reaching the same contact position as defined by independent quasi-static experiments. We suspect the dry contact is made at asperities while fluid can persist in the valley as a result of roughness, but don't verify it specifically. In addition, air bubbles are unlikely to nucleate because of the wettability of the silicone oil on both opposing surfaces (see Figure S3 in supporting information). We observe that, for $T = \delta/\sqrt{Rh_0} \sim 0.3$, two small pockets of fluid are still visible in the 2D images when contact is made (blue arrows in Fig. 6c). Therefore, from a 3-dimensional view, a continuous ring of liquid is trapped when the edges of surfaces touch each other. This complex contact morphology is a direct consequence of elasticity and coating thickness, while the effect of apparent slip (due to roughness) allows the surfaces to make contact. We also plot the contact shape of a much thicker film ($T = \delta/\sqrt{Rh_0} \sim 1.6$, Fig. 6b), at the time when the edges of the dimple touch and makes contact with the opposing surface. Unlike the case of $\delta/\sqrt{Rh_0} \sim 0.3$, a large and continuous volume of fluid is trapped at the center, showing that the coating thickness has a significant effect on contact morphology, but that roughness is still necessary to trap fluid in contact. In the absence of roughness, the deformation would relax prior to contact at a significantly later time. There is an important distinction in the dynamic of contact formation of soft solids compared to fluid droplets. For droplets, when a dimple or wimple forms a jump-in and coalescence often happens at contact points. For the case of soft solids, the fluid trapped will have drain slowly through asperities. We show that the contact points are determined by the elasticity and coating thickness, while the time required for contact is largely accelerated by surface roughness.

## Conclusions

In summary, we present the first experimental observation on effects of apparent surface slip caused by roughness on the morphology and dynamics of soft contacts. We disentangle the contributions of surface stratification and slip through measurement of absolute fluid film thickness and comparing with theory based on linear elasticity and fluid lubrication. We show that stratification leads to deformation profiles that are drastically different from those observed for elastic half-space (wimples). The coupled effect of apparent surface slip and elasticity can lead to trapped fluid pockets between surfaces in contact. The contact morphology between soft coatings has important implications in tribology[67], interfacial rheology[68], pressure sensitive adhesives[69], and in biology[70]. In the context of bio-inspired adhesives, the surface structure and mechanical properties of soft coatings are engineered to maximize the contributions of surface forces (such as van der Waals interactions)[71]. Here we see that the dynamics of contact formation, and associated elastic deformation, can prevent physical contact therefore could play an important role in the adhesive performance of bio-inspired adhesives. Similarly, trapped fluid during the bonding of pressure sensitive adhesives could lead to poor bond formation and weaker performance. Lastly, with the absence of surface profile information, the interplay between stratification and roughness could lead to misinterpretation of dynamic surface forces measurements.

# Acknowledgments

We acknowledge Prof. Patricia McGuiggan for her assistance on AFM measurements. We also thank Matthew Ryan Tan for his help on algorithm development, and Paul Roberts for indentation measurements. This work was supported by NSF-CMMI 1538003. Y.W. acknowledge the support from the Science Fundamental of China University of Petroleum, Beijing (No.2462018YJRC002).

# References

1. D. Dowson, *Wear*, 1995, **190**, 125-138.
2. K. Sathyan, H. Y. Hsu, S.-H. Lee and K. Gopinath, *Tribol. Int.*, 2010, **43**, 259-267.
3. P. Lugt and G. E. Morales-Espejel, *Tribol. Trans.*, 2011, **54**, 470-496.
4. K. Nanjundiah, P. Y. Hsu and A. Dhinojwala, *J. Chem. Phys.*, 2009, **130**, 024702.
5. B. Persson, *J. Phys.: Condens. Matter*, 2007, **19**, 376110.
6. J. Iturri, L. Xue, M. Kappl, L. García-Fernández, W. J. P. Barnes, H. J. Butt and A. del Campo, *Adv. Funct. Mater.*, 2015.
7. J. Li, A. Celiz, J. Yang, Q. Yang, I. Wamala, W. Whyte, B. Seo, N. Vasilyev, J. Vlassak and Z. Suo, *Science*, 2017, **357**, 378-381.
8. S. Jahn, J. Seror and J. Klein, *Annu. Rev. Biomed. Eng.*, 2016, **18**, 235-258.
9. D. M. Drotlef, L. Stepien, M. Kappl, W. J. P. Barnes, H. J. Butt and A. del Campo, *Adv. Funct. Mater.*, 2013, **23**, 1137-1146.
10. W. Federle, W. Barnes, W. Baumgartner, P. Drechsler and J. Smith, *J. R. Soc. Interface*, 2006, **3**, 689-697.
11. H. Yuk, S. Lin, C. Ma, M. Takaffoli, N. X. Fang and X. Zhao, *Nat. Commun.*, 2017, **8**, 14230.
12. J.-H. Dirks and W. Federle, *Soft Matter*, 2011, **7**, 11047-11053.

13. P. Huber, *J. Phys.: Condens. Matter*, 2015, **27**, 103102.
14. R. M. Espinosa-Marzal, R. M. Bielecki and N. D. Spencer, *Soft Matter*, 2013, **9**, 10572-10585.
15. M. Scaraggi, G. Carbone, B. N. J. Persson and D. Dini, *Soft Matter*, 2011, **7**, 10395-10406.
16. S. Leroy, A. Steinberger, C. Cottin-Bizonne, F. Restagno, L. Leger and E. Charlaix, *Phys. Rev. Lett.*, 2012, **108**, 264501.
17. T. Salez and L. Mahadevan, *J. Fluid Mech.*, 2015, **779**, 181-196.
18. T. J. Baudouin Saintyves, Thomas Salez, and L. Mahadevan, *Proc. Natl. Acad. Sci.*, 2016, **113**, 5847-5849.
19. H. Davies, D. Débarre, C. Verdier, R. P. Richter and L. Bureau, *arXiv preprint arXiv:1711.11378*, 2017.
20. Z. Jin, D. Dowson and J. Fisher, *Proceedings of the Institution of Mechanical Engineers, Part H: Journal of Engineering in Medicine*, 1992, **206**, 117-124.
21. B. K. Ahn, S. Das, R. Linstadt, Y. Kaufman, N. R. Martinez-Rodriguez, R. Mirshafian, E. Kesselman, Y. Talmon, B. H. Lipshutz and J. N. Israelachvili, *Nat. Commun.*, 2015, **6**, 8663.
22. Y. Wang, G. A. Pilkington, C. Dhong and J. Frechette, *Curr. Opin. Colloid Interface Sci.*, 2017, **27**, 43-49.
23. E. Charrault, T. Lee, C. D. Easton and C. Neto, *Soft Matter*, 2016.
24. R. Lhermerout and S. Perkin, *Phys. Rev. Fluids*, 2018, **3**, 014201.
25. C. Buchcic, R. Tromp, M. Meinders and M. C. Stuart, *Soft Matter*, 2017.
26. C. Creton and M. Ciccotti, *Rep. Prog. Phys.*, 2016, **79**, 046601.
27. S. Yashima, N. Takase, T. Kurokawa and J. P. Gong, *Soft Matter*, 2014, **10**, 3192.
28. D. Tabor and R. H. S. Winterton, *Proc. R. Soc. London, Ser. A*, 1969, **312**, 435-450.
29. J. N. Israelachvili, *J. Colloid Interface Sci.*, 1973, **44**, 259-272.
30. J. N. Israelachvili, Y. Min, M. Akbulut, A. Alig, G. Carver, G. W. Greene, K. Kristiansen, E. E. Meyer, N. S. Pesika and K. J. Rosenberg, *Rep. Prog. Phys.*, 2010, **73**, 036601-036601−036616.
31. E. Blomberg, P. M. Claesson and H. K. Christenson, *J. Colloid Interface Sci.*, 1990, **138**, 291-293.
32. N. Bowden, S. Brittain, A. G. Evans, J. W. Hutchinson and G. M. Whitesides, *Nature*, 1998, **393**, 146-149.
33. S. Tolansky, *Multiple-Beam Interferometry of surfaces and films*, Oxford University Press, London, 1948.
34. R. Gupta and J. Fréchette, *J. Colloid Interface Sci.*, 2013, **412**, 82-88.
35. P. Roberts, G. A. Pilkington, Y. Wang and J. Frechette, *Rev. Sci. Instrum.*, 2018, **89**, 043902.
36. Y. Wang, M. R. Tan and J. Frechette, *Soft Matter*, 2017, **13**, 6718.
37. O. I. Vinogradova and G. E. Yakubov, *Phys. Rev. E* 2006, **73** 045302.
38. O. Reynolds, *Proc. R. Soc. Lond.*, 1886, **40**, 191-203.
39. R. H. Davis, J.-M. Serayssol and E. Hinch, *J. Fluid Mech.*, 1986, **163**, 479-497.
40. K. R. Shull, *Mater. Sci. Eng. R-Rep.*, 2002, **36**, 1-45.
41. C. Pick, C. Argento, G. Drazer and J. Frechette, *Langmuir*, 2015, **31**, 10725-10733.
42. Y. Wang, Johns Hopkins University, 2017.
43. S. Leroy and E. Charlaix, *J. Fluid Mech.*, 2011, **674**, 389-407.
44. N. Balmforth, C. Cawthorn and R. Craster, *J. Fluid Mech.*, 2010, **646**, 339-361.
45. J. Li and T.-W. Chou, *Int .J. Solids. Struct.*. 1997, **34**, 4463-4478.
46. T. Nogi and T. Kato, *J. Tribol.*, 1997, **119**, 493-500.
47. E. Gacoin, C. Fretigny, A. Chateauminois, A. Perriot and E. Barthel, *Tribol. Lett.*, 2006, **21**, 245.
48. A. Perriot and E. Barthel, *J. Mater. Res.*, 2004, **19**, 600-608.
49. F. K. Yang, W. Zhang, Y. G. Han, S. Yoffe, Y. C. Cho and B. X. Zhao, *Langmuir*, 2012, **28**, 9562-9572.
50. L. Y. Clasohm, J. N. Connor, O. I. Vinogradova and R. G. Horn, *Langmuir*, 2005, **21**, 8243-8249.
51. Y. Wang, C. Dhong and J. Frechette, *Phys. Rev. Lett.*, 2015, **115**, 248302.
52. Z. A. Levine, M. V. Rapp, W. Wei, R. G. Mullen, C. Wu, G. H. Zerze, J. Mittal, J. H. Waite, J. N. Israelachvili and J.-E. Shea, Proc. Natl. Acad. Sci., 2016, **113**, 4332-4337.

53. R. Manica, J. N. Connor, S. L. Carnie, R. G. Horn and D. Y. Chan, *Langmuir*, 2007, **23**, 626-637.
54. V. Popov, *Contact mechanics and friction: physical principles and applications*, Springer Science & Business Media, 2010.
55. C. Neto, D. R. Evans, E. Bonaccurso, H.-J. Butt and V. S. Craig, *Rep. Prog. Phys.*, 2005, **68**, 2859.
56. R. Gupta and J. Frechette, *Langmuir*, 2012, **28**, 14703-14712.
57. F. Kaveh, J. Ally, M. Kappl and H.-J. r. Butt, *Langmuir*, 2014, **30**, 11619-11624.
58. C. D. Honig and W. A. Ducker, *Phys. Rev. Lett.*, 2007, **98**, 028305.
59. S. Richardson, *J. Fluid Mech.*, 1973, **59**, 707-719.
60. O. I. Vinogradova and F. Feuillebois, *J. Colloid Interface Sci.*, 2000, **221**, 1-12.
61. E. Bonaccurso, H.-J. Butt and V. S. Craig, *Phys. Rev. Lett.*, 2003, **90**, 144501.
62. G. A. Pilkington, R. Gupta and J. Fréchette, *Langmuir*, 2016, **32**, 2360-2368.
63. M. Stefan, *Akad. Wiss. Wien (Abt. II Math. Phys.)*, 1874, **69**, 713.
64. D. Y. Chan, E. Klaseboer and R. Manica, *Soft Matter*, 2011, **7**, 2235-2264.
65. C. Kunert, J. Harting and O. I. Vinogradova, *Phys. Rev. Lett.*, 2010, **105**, 016001.
66. A. Mongruel, T. Chastel, E. S. Asmolov and O. I. Vinogradova, *Phys. Rev. E*, 2013, **87**, 011002.
67. J. Snoeijer, J. Eggers and C. Venner, *Physics of Fluids (1994-present)*, 2013, **25**, 101705.
68. M. Gross, T. Kruger and F. Varnik, *Soft Matter*, 2014, **10**, 4360-4372.
69. P. Karnal, P. Roberts, S. Gryska, C. King, C. Barrios and J. Frechette, *ACS Appl. Mater. Interfaces*, 2017, **9**, 42344-42353.
70. A. Gopinath and L. Mahadevan, *Proc. R. Soc. London, Ser. A*, 2011, rspa20100228.
71. M. D. Bartlett, A. B. Croll, D. R. King, B. M. Paret, D. J. Irschick and A. J. Crosby, *Adv. Mater.*, 2012, **24**, 1078-1083.

Supporting information for

# Morphology of soft and slippery contact via fluid drainage

Yumo Wang[1,3] and Joelle Frechette[1,2]

[1]Chemical and Biomolecular Engineering Department and [2]Hopkins Extreme Materials Institute, Johns Hopkins University, Baltimore MD 21218 USA. \

[3]National Engineering Laboratory for Pipeline Safety, Beijing Key Laboratory of Urban Oil and Gas Distribution Technology, China University of Petroleum, Beijing, 18# Fuxue Road, Changping District, 102249 Beijing, China

(Dated Aug 30th, 2018)

## 1. Summary of experimental parameters

| Type | Physical parameter | Sample 1 | Sample 2 | Sample 3 | Sample 4 |
|---|---|---|---|---|---|
| **PDMS film** | Thickness, $T$ | 330 μm | 147 μm | 54 μm | 11 μm |
| | Disk 1: $R_1$ | 1.62 cm | 1.32 cm | 1.64 cm | 1.18 cm |
| | Disk 2: $R_2$ | 1.89 cm | 1.86 cm | 2.28 cm | 1.02 cm |
| | $R_h = 2R_1R_2/(R_1+R_2)$ | 1.74 cm | 1.53 cm | 1.91 cm | 1.09 cm |
| | $R_g = R = (R_1R_2)^{1/2}$ | 1.75 cm | 1.55 cm | 1.91 cm | 1.10 cm |
| | Young's modulus, $E$ | 0.95 MPa | 1.05 MPa | 1.05 MPa | 0.95 MPa |
| | Poisson's ratio, $v$ | 0.5 | | | |
| **SFA** | Spring constant, $k$ | 165.3 N/m | | | |
| | Drive velocity, $V$ | 50-150 nm/s | | | |
| | Initial separation, $h(0,0)$ | 2-5 μm | | | |
| | Maximum motor travel | 10 μm | | | |
| | SU-8 thickness | 6-7 um | | | |
| | Mica thickness | 3-10 um | | | |
| | Top silver thickness | 50 nm | | | |
| **Fluid (silicone oil)** | Viscosity, $\eta_{fluid}$ | 0.2 Pa·s | | | |
| | Density, $\rho$ | 0.98 g/cm$^3$ | | | |

**Table S1:** Overview of the experimental parameters investigated.

Our experiments are performed between crossed-cylinders (equivalent to the sphere-plane geometry when R >> h) using the Surface Forces Apparatus (SFA). One surface is rigid (bottom in Fig. S1) and the other is compliant due to the presence of a PDMS film (polydimethylsiloxane) coated with a 50 nm silver film as a top layer and deposited, silver side up, on a mica sheet (Fig. S1). The silver layer serves a mirror for the interferometer and prevents swelling of the PDMS with the silicone oil fluid.

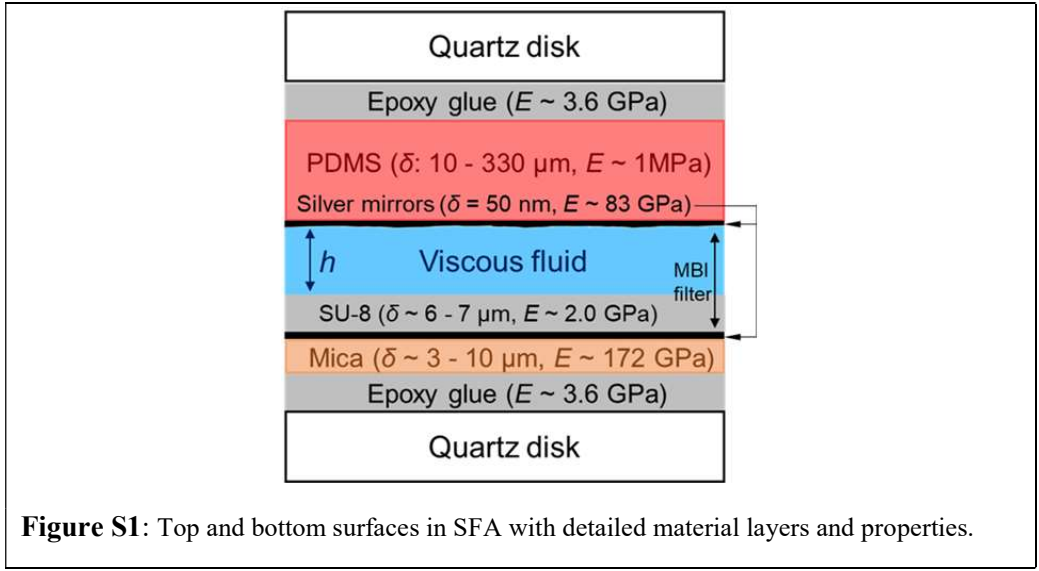

**Figure S1**: Top and bottom surfaces in SFA with detailed material layers and properties.

## 2. Surface characterization

### 2.1. Determination of Young's modulus of the bulk PDMS

We determine independently the Young's modulus of polydimethyl siloxane (PDMS) via indentation measurements with a spherical probe. We use a homemade Multimode Force Microscope[1] and compare the force-indentation results with layered indentation theory. The contact area is measured by a high-speed CMOS camera (FLIR GS3-U3-23S6C-C) and fitted to circular shape using ImageJ to determine contact radius. Meanwhile, the normal applied force is measured by a cantilever spring ($k$=1055$N/m$), and the radius of the indentor is 3.175 $mm$, which is greater than the coating thickness.

To obtain the Young's modulus of PDMS we need to account for the effect of stratification on the indentation measurements. We follow the analytical corrections presented by Shull[2], applying corrections factors on indentation depth ($\sigma$), normal force ($F$) and contact radius ($a$) to a Hertzian contact model. The correction factors used are listed follows:

$$f_p(a/\delta) = (1 + \beta(a/\delta)^3) \tag{S1}$$

$$f_\sigma(a/\delta) = \left(0.4 + 0.6\exp\left(\frac{-1.8a}{\delta}\right)\right) \tag{S2}$$

$$\frac{1}{f_c(a/\delta)} = 1 + \left(\frac{0.75}{(a/\delta) + (a/\delta)^3} + \frac{2.8(1-2\nu)}{a/\delta}\right)^{-1}, \tag{S3}$$

and the Hertzian contact model changes to:

$$\sigma = \frac{a^2}{R} f_\sigma(a/\delta) + \frac{f_c(a/\delta)}{2E^*\delta}\left(F - \frac{4E^*a^3}{3R}f_p(a/\delta)\right), \tag{S4}$$

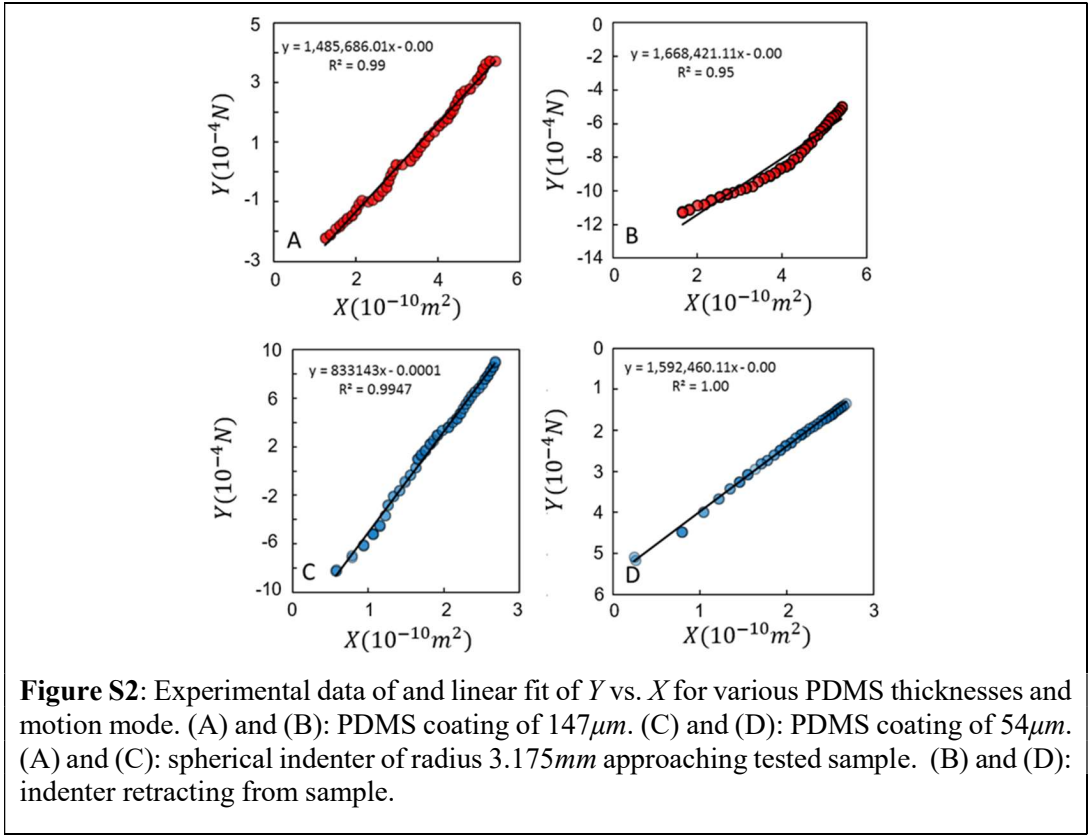

**Figure S2**: Experimental data of and linear fit of $Y$ vs. $X$ for various PDMS thicknesses and motion mode. (A) and (B): PDMS coating of $147\mu m$. (C) and (D): PDMS coating of $54\mu m$. (A) and (C): spherical indenter of radius $3.175mm$ approaching tested sample. (B) and (D): indenter retracting from sample.

which can be rearranged to:

$$\left(\left(\delta - \frac{a^2}{R}f_\sigma(a/h)\right)2a + f_c(a/h)\frac{4a^3}{3R}f_p(a/h)\right)E^* = f_c(a/h)F \;, \tag{S5}$$

or in simplified notations:

$$XE^* = Y \;. \tag{S6}$$

The experimental data are then compared with the linear relationship between $Y$ and $X$ to extract $E^*$ as the slope. The results are shown in Figure S2 for PDMS films with two thicknesses ($147\mu m$, A and B; $54\mu m$, C and D), for both approach (A and C) and retraction (B and D). The plots show good general linearity in $Y$ vs. $X$, which verifies the suitability of the corrections used. The average measured reduced Young's modulus of PDMS is then calculated as $1.39MPa$ and Young's modulus $E = E^*(1-v^2)$ as $1.05\pm0.15MPa$ with Poisson's ratio set to be 0.5. We also performed indentation experiments directly in the SFA in our previous work[3], where we treated a $330\mu m$ thick film as an approximated half-space and we applied semi-infinite JKR theory to get a modulus of $1.08\pm0.05MPa$, including the rigid substrate contributions. If we re-fit same data with layered equations mentioned above, the absolute modulus without the substrate effects obtained is $0.95\pm0.05MPa$. The difference between the two measurements is therefore about 9.1%. To compare our results with different thicknesses, we apply here the $0.95MPa$ modulus to the $330\mu m$ film.

## 2.2. Characterization of the silvered-PDMS surfaces

Plasma-treatment of PDMS leads to a surface roughness on the order of 10$nm$ [4]. In addition, subsequent thermal evaporation of silver on PDMS can further increase the surface roughness through the formation of small cracks of the oxidized surface layer, although large scale buckling is generally prevented [5]. We characterized the roughness of the surfaces employed in the SFA using an atomic force microscope, AFM (Dimension 3100, Bruker Nano, CA). We characterized the roughness (peak-to-valley difference) on silver-coated PDMS films with PDMS thickness of $\delta = 54\,\mu m$ and $\delta = 147\,\mu m$ and a silver thickness of 50 $nm$. Imaging was performed in the contact mode, on an area of $3 \times 3\,\mu m^2$ at a scan rate of $1.50 Hz$ (Figure S3A, B). We found that within a sample (PDMS thickness) the root-mean square (RMS) roughness and peak-to-valley difference was fairly constant but that it varied from sample to sample. The $\delta = 54\,\mu m$ sample has an average (RMS) roughness of 3.33$nm$, and a peak-to-valley difference of 22.00$nm$, while $\delta = 147\,\mu m$ film has a RMS roughness of 5.8 $nm$ and a peak-to-valley difference of 49.9$nm$.

Poor wetting may generate surface slip and nano-bubbles during drainage. We measured the contact angle of silicone oil on top of silvered-PDMS surface, as shown in Figure S3C, and the average contact angle is 16.5 degrees. As expected, and verified here, the silicone oil wets the silvered-PDMS.

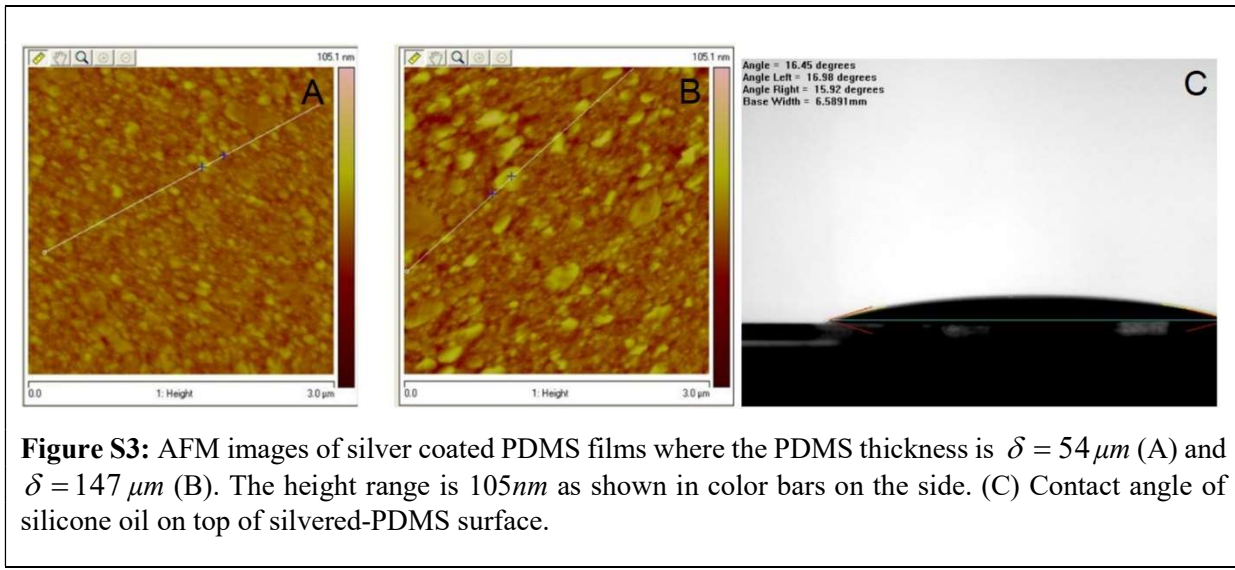

**Figure S3:** AFM images of silver coated PDMS films where the PDMS thickness is $\delta = 54\,\mu m$ (A) and $\delta = 147\,\mu m$ (B). The height range is 105$nm$ as shown in color bars on the side. (C) Contact angle of silicone oil on top of silvered-PDMS surface.

## 2.3. Predictions on surface morphology with different thickness and roughness

The apparent slip and elasticity have opposing but qualitatively different role on the drainage process. The morphology of the surface is almost uniquely determined by the mechanical properties of surfaces (coating thickness and elasticity), while the fluid gaps shifts due to the apparent slip caused by roughness. To clarify this point we show here in Figure S4 how varying the apparent slip length will change the fluid film thickness but not the morphology, as observed in our experiments.

The surface morphology using single shifted-plane value of 35$nm$ is showed by blue solid line in Figure S4. We can see that: the experimental data agree mostly with the prediction with a slight underestimation. Overall, the shape of the predicted curve recovers very well the wimple shape as observed during experiment. The shifted length (representing level of roughness) is varied in dotted and dot-dashed blue lines, with a value of 15 and 45$nm$ used, respectively. Besides expected deviations on surface separation for 15 and 45$nm$ shifted length, the wimple shape is mostly recovered on these two curves, indicating that the surface morphology is uniquely determined by mechanical properties, while the absolute surface position is highly influenced by surface roughness during drainage. This effect is further proved by the green solid line and black dotted line, in which incorrect coating thicknesses are used. For the green solid line, an underestimated thickness is used ($\delta = 30\mu m$) while a no-slip boundary condition is incorporated to compensate the deviation on central position. We can see that: the morphology predicted in this case is clearly off, especially at the central regime. On the contrary, if we assign an overestimated coating thickness (half-space, black dotted line). In this case, even if we compensate the central separation by using a large shifted length of 110$nm$, the morphology of experimental results cannot be acquired, especially for the radial position > 100$\mu m$, as a significant underestimation is observed. These analyses prove that although both surface stratification and roughness facilitate drainage, their role played are qualitatively different, and in fact distinguishable from and only from the surface morphology. In our SFA experiments, the coating thickness and material Young's modulus are independently measured, therefore, the morphology of the soft coating is predicted without fitting parameter, and the shifted length is the only fitting parameter to predict the absolute separation. Thus, the role played by mechanical properties and surface roughness are decoupled.

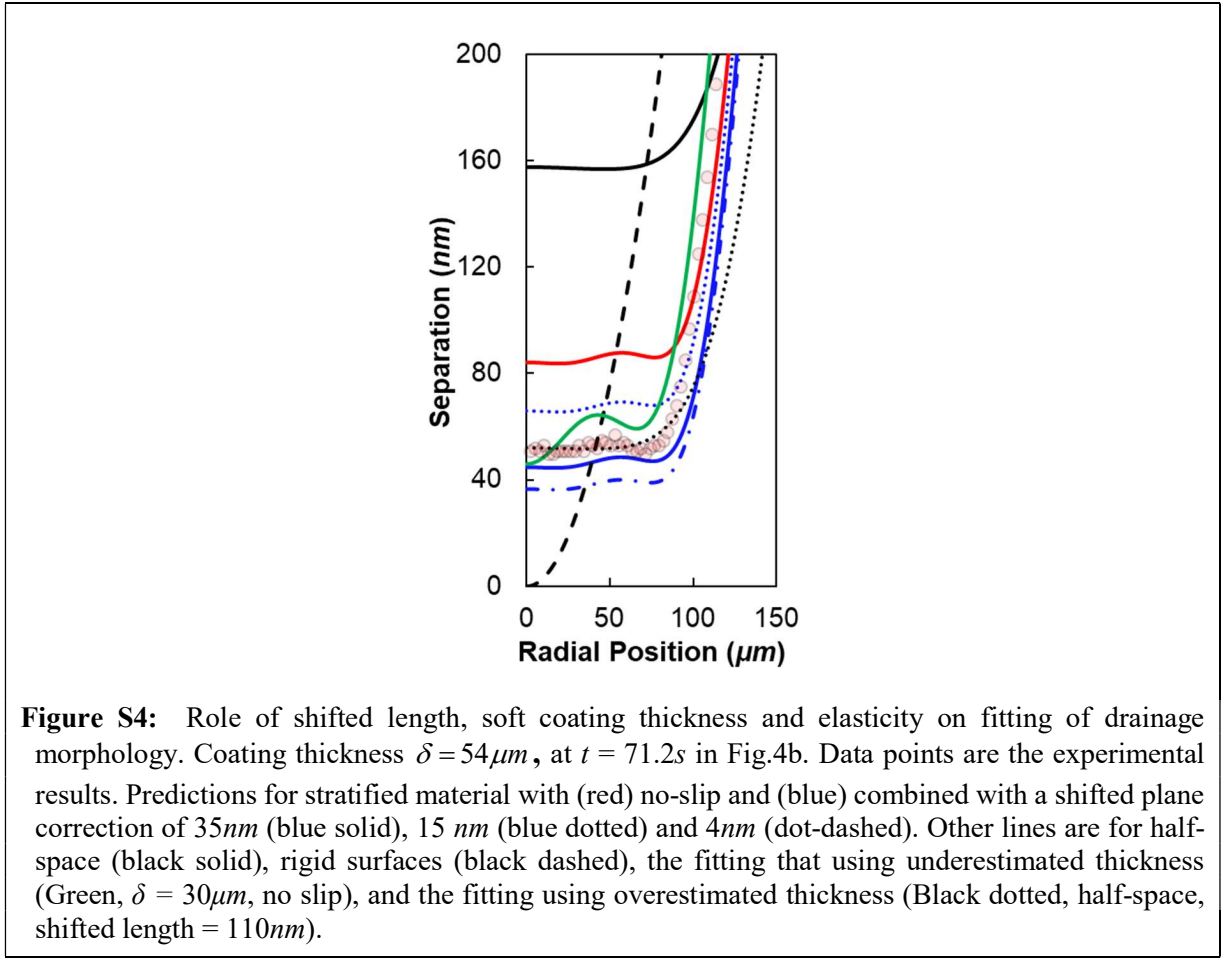

**Figure S4:** Role of shifted length, soft coating thickness and elasticity on fitting of drainage morphology. Coating thickness $\delta = 54\mu m$, at $t = 71.2s$ in Fig.4b. Data points are the experimental results. Predictions for stratified material with (red) no-slip and (blue) combined with a shifted plane correction of 35$nm$ (blue solid), 15 $nm$ (blue dotted) and 4$nm$ (dot-dashed). Other lines are for half-space (black solid), rigid surfaces (black dashed), the fitting that using underestimated thickness (Green, $\delta = 30\mu m$, no slip), and the fitting using overestimated thickness (Black dotted, half-space, shifted length = 110$nm$).

### 2.4. Effects of the thin silver layer (50 nm) on top of the soft layer

We do not expect the 50$nm$ silver film on the PDMS to have significant constraint on the deformation of PDMS layer, because of the thickness of the silver film compared to that of the PDMS (50$nm$ vs 10-330 microns) along with the large radius of curvature and hydrodynamic radius in the SFA(1-2$cm$). In addition, the silver film has the added advantage of preventing swelling of the PDMS in the oil by blocking transport of the oil into the PDMS. It is possible that the top silver layer constrains the deformation of the PDMS layer, and could have an effect on the Young's modulus. Therefore, to estimate the role played by the silver film, a quick and convenient method to verify its contribution to an effective modulus is to use the numerical results for indentation of stratified films provided in Figures 2 & 3 of Ref. [6]. The Young's modulus of silver layer is at least 1000 times larger than the PDMS base (even for thin PDMS coating in our experiments, in which $\delta = 10.9$ $\mu m$, the effective modulus in this case is ~ 80$MPa$, as verified by indentation experiment), and the hydrodynamic radius much greater than the silver film thickness, the increase in Young's modulus result from the silver coating should be very small.

A more quantitative method to estimate the stiffening effects due to silver coating is to use the approximation of Eqn. 11 in Ref. [7]:

$$F_f = \pi E_f^* Y \left[ \frac{-6a^6 + 6a^4(R^+)^2 + Y^2 a^4 - 4Y^2 a^2 (R^+)^2}{6((R^+)^2 - a^2)^{5/2}} \right]. \quad (S7)$$

In which, $F_f$ is the effective extra load due to metal layer, $E_f^*$ is the effective modulus of the metal film, $R^+$ is the curvature of the deformation during indentation. $Y$ is the thickness of the metal film on top. We can further compare the modulus of the PDMS-Silver layers with and without the silver to get:

$$\frac{E_{sys(l)}^*}{E_{sys}^*} = 1 + \frac{3R\pi Y}{4a^3} \left[ \frac{-6a^6 + 6a^4(R^+)^2 + Y^2 a^4 - 4Y^2 a^2 (R^+)^2}{6((R^+)^2 - a^2)^{5/2}} \right]. \quad (S8)$$

Based on Eqn. S8, we estimate the actual importance of silver layer on the apparent Young's modulus. The results show that the influence of the silver layer increases with the contacting area. At the largest possible hydrodynamic radius in our experiment (~100$s$ of $\mu m$), the stiffening due to the silver film is only about 0.5%, we can safely neglect the effect of the silver layer in our application.

## 3. SFA data processing and analysis

### 3.1. Multiple beam interferometry (MBI)

the SFA is designed to measure the forces acting between two surfaces as a function of their separation. The force is measured via the deflection of a cantilever on which the lower surfaces is mounted, and the surface separation is deduced independently using multiple beam interferometry (MBI)[8]. MBI relies on the construction of an interference filter, which consists of two reflective surfaces (here silver) separated by a dielectric medium. When white light is passed through such a filter, only certain wavelengths will emerge; all the other wavelengths interfere destructively and are not transmitted. The emerging wavelengths result in observable fringes (fringes of equal chromatic order, FECO). Optical theory relates the transmitted wavelength to an absolute surface separation with a resolution of 1-3 $Å$. The resolution in normalized force is 1$\mu N/m$. The separation between the surfaces is varied by using two microstepping motors moving the upper and lower micrometer rods. Two 50 $nm$-thick silver films, being highly reflective (>95%), yield sharp and high-contrast primary FECO. We use a CCD camera: Q-Imaging Retiga 4000RV, to capture the digital image of such FECO images after they are dispersed in an imaging spectrograph. Some examples of original FECO images that's appear on the spectrograph is shown below in Figure S5. Some of the snapshot of FECO images are colorized using ImageJ and shown directly in Figure 2 and Figure 6 to reveal contacting shape. For acquiring quantitative absolute separation, further analysis is needed as we present in next paragraph.

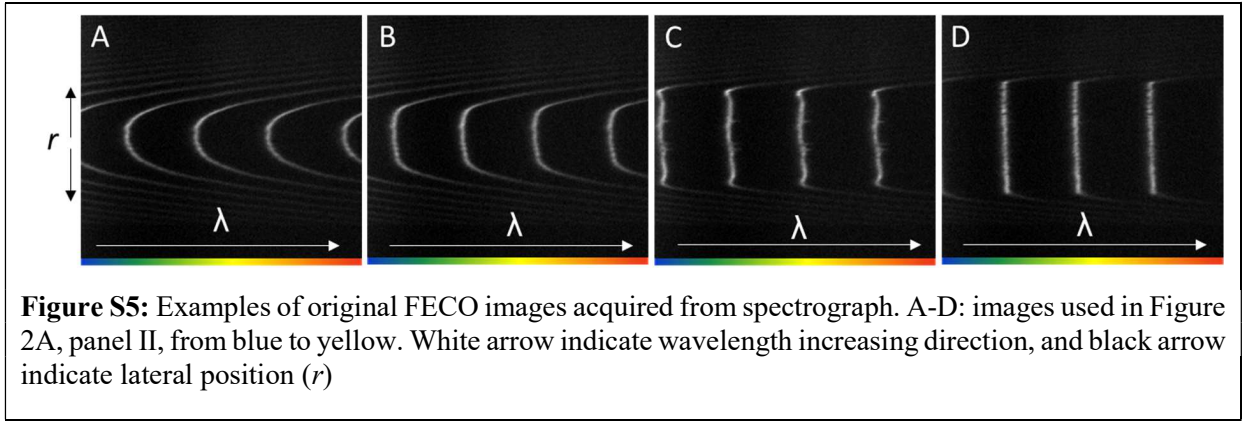

**Figure S5:** Examples of original FECO images acquired from spectrograph. A-D: images used in Figure 2A, panel II, from blue to yellow. White arrow indicate wavelength increasing direction, and black arrow indicate lateral position ($r$)

### 3.2. Data analysis

The center of mass (CoM) of the measured intensity as a function of pixel position is used to estimate pixel position for each of the peaks in the intensity. We use the pixel positions of the known discrete spectral wavelengths (green: 546*nm* and orange-yellow doublet: 577-579*nm*) of mercury to convert pixels into wavelengths of the FECO. Upon gluing to the cylindrical lens, the interference filter acquires curvature and the medium thickness and consequently, the wavelengths of the FECO are a function of lateral position. The presence of macroscopic surfaces enables the visualization of the 2-dimensional geometry of the interacting surfaces, which is reflected in the shape of the FECO.

The FECO obtained for two crossed-cylinders are used to calculate the radii of curvature of the interacting cylinders (Section 1). The wavelengths at the vertex of the parabolic fringes are used to estimate the surface separation at the point of closet approach (PCA) for a sphere-plane configuration. Estimation of radii of curvature and diameter of zone of contact requires the lateral resolution (typically ≈ 2.64*μm* as dictated by the optical magnification). The use of highly reflective silver suppresses the secondary, tertiary, and gap fringes. Note here that we use the center of mass of the interferometric fringes to determine their wavelengths and calculate the fluid film thickness. The FECO represent the summation of all intensity profiles corresponding to the point of lights transmitted through silver mirrors [9]. As a result, the fluid film thickness we measured is based on the center of mass of the asperities within each pixel (~2*μm*).

We employ the fast spectral correlation (FSC) algorithm devised by Heuberger to facilitate transformation or inversion of transmitted intensity to unknown medium thickness and/or index of refraction [10]. In FSC, the sum of transmitted intensity for multiple experimentally observed wavelengths for a range of medium thickness and/or index of refraction is examined, and the unknown parameter or combination of parameters that results in the maximum transmissivity is used as the solution. We use an in-house LabVIEW interface to facilitate dynamic measurements as well as automated analysis of FECO images. For more details regarding the image processing and analysis, the readers are directed to Ref. [11].

## 4. Definition of contact position (zero separation)

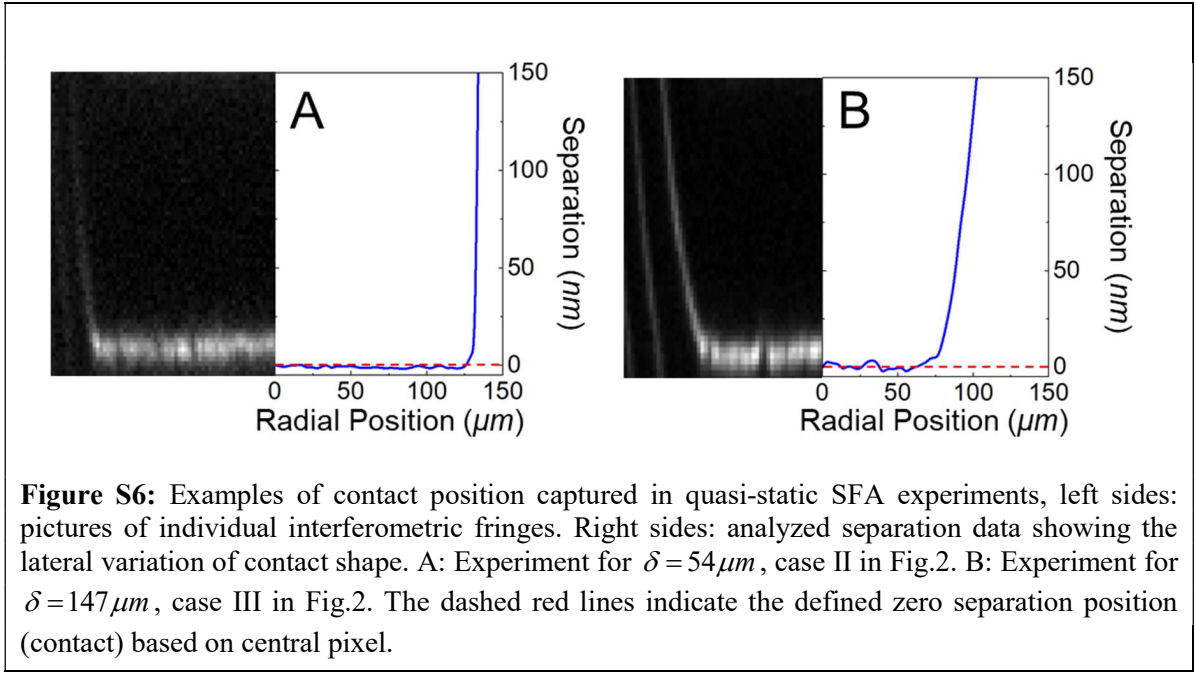

**Figure S6:** Examples of contact position captured in quasi-static SFA experiments, left sides: pictures of individual interferometric fringes. Right sides: analyzed separation data showing the lateral variation of contact shape. A: Experiment for $\delta = 54\,\mu m$, case II in Fig.2. B: Experiment for $\delta = 147\,\mu m$, case III in Fig.2. The dashed red lines indicate the defined zero separation position (contact) based on central pixel.

The contact position (zero separation) is defined through quasi-static experiments in which the surfaces were slowly pushed together until a large central contact regime is formed (Fig. S6). We mark the position at the center pixel (point of closest approach) and use it as the reference position for zero separation throughout (red dashed lines in Fig. S6). The lateral variation of this contact position is typically less than 0.5 *nm*, as shown in Fig.S6A for case II in Fig.2, and same level of smoothness applies to Case I and IV in Fig.2. The lateral variation of contact position in these cases can be attributed to the roughness of the silver layer. For some cases (case III in Fig. 2) the tilting or unevenness of underlying SU-8 can affect the contact morphology in which the maximum spacial variation can be 5 *nm*, as shown by Fig. S6B.

**References**

[1] P. Roberts, G.A. Pilkington, Y. Wang, J. Frechette, A multifunctional force microscope for soft matter with in situ imaging, Review of Scientific Instruments, 89 (2018) 043902.

[2] K.R. Shull, Contact mechanics and the adhesion of soft solids, *Materials Science and Engineering: R: Reports* 36 (2002) 1-45.

[3] Y. Wang, C. Dhong, J. Frechette, Out-of-contact elastohydrodynamic deformation due to lubrication forces, *Physical Review Letters*, 115 (2015) 248302.

[4] J. Bongaerts, K. Fourtouni, J. Stokes, Soft-tribology: lubrication in a compliant PDMS–PDMS contact, *Tribology International* 40 (2007) 1531-1542.

[5] J.P. Déry, D. Brousseau, M. Rochette, E.F. Borra, A.M. Ritcey, Aluminum-coated elastomer thin films for the fabrication of a ferrofluidic deformable mirror, *Journal of Applied Polymer Science*, 134 (2017) 44542.

[6] A. Perriota and E. Barthel, Elastic contact to a coated half-space: Effective elastic modulus and real penetration, *Journal of Materials Research,* 19.2 (2004): 600-608.

[7] F. K. Yang, W. Zhang, Y. G. Han, S. Yoffe, Y. C. Cho, and B. X. Zhao, "Contact" of Nanoscale Stiff Films, *Langmuir* 28, 9562 (2012).

[8] J. Israelachvili, Thin film studies using multiple-beam interferometry, *Journal of Colloid & Interface Science*, 44 (1973) 259-272.

[9] J.M. Levins, T.K. Vanderlick, Impact of roughness of reflective films on the application of multiple beam interferometry, *Journal of Colloid & Interface Science*, 158 (1993) 223-227.

[10] M. Heuberger, The extended surface forces apparatus. Part I. Fast spectral correlation interferometry, *Review of Scientific Instruments*, 72 (2001) 1700-1707.

[11] Y. Wang, Deformation of compliant coatings due to lubrication forces, Ph.D. thesis, Johns Hopkins University, 2017.